\documentclass[
superscriptaddress,
preprint,
longbibliography,
amsmath,
amssymb,
prab,
aps,
floatfix,
]{revtex4-2}
\usepackage{graphicx}
\usepackage{bm}
\usepackage{tikz,xcolor,hyperref}
\usepackage{microtype}
\hypersetup{
  hidelinks,
  colorlinks   = true, 
  urlcolor     = blue, 
  linkcolor    = blue, 
  citecolor   = blue 
}
\DeclareMathOperator{\sinc}{sinc}

\date{\today}
\begin{document}

\title{Simulation of longitudinal   
Landau damping in bunches with space charge
}

\author{Oliver Boine-Frankenheim}
\affiliation{Technische Universität Darmstadt, Schlossgartenstraße 8, 64289 Darmstadt, Germany}
\affiliation{GSI Helmholtzzentrum für Schwerionenforschung GmbH, Planckstraße 1, 64291 Darmstadt, Germany}

\author{Thilo Egenolf}
\affiliation{Technische Universität Darmstadt, Schlossgartenstraße 8, 64289 Darmstadt, Germany}

\begin{abstract}
For a single hadron bunch affected by longitudinal space charge in a stationary rf bucket we analyze the frequency spectrum close to the expected loss of Landau damping for the lowest order dipole mode. For different bunch intensity parameters we obtain the bunch oscillation spectrum from a conventional longitudinal particle tracking code with a grid-based space charge solver. We validate selected results against a grid-less space charge solver. We highlight the importance of the choice of the cut-off parameter $h_c$ in the space charge impedance for the long-term accuracy of grid-based schemes. For typical bunch parameters in an ion synchrotron at injection energies we find that the branching point, where the dipole mode frequency emerges from the incoherent synchrotron frequency spectrum, as well as the damping of the dipole mode do not depend on $h_c$, chosen well below the actual value for realistic beam pipes.  
\end{abstract}

\maketitle

\section{Introduction}

The prediction of the intensity threshold for the loss of Landau damping in hadron bunches with space charge and other broadband impedances in synchrotrons and colliders has attracted much interest in the last decades. Below transition energy and for 'regular' matched bunch distributions, space charge alone does not lead to instabilities, but it can alter Landau damping. Therefore, accurate predictions of the thresholds for the loss of Landau damping are helpful to determine the need for active bunch stabilization measures, for example.   

For the rigid bunch dipole mode and a local elliptic bunch distribution, threshold space charge parameters were predicted in \cite{Hofmann1979a}, based on simple analytic criteria. Equating the analytically known rigid bunch oscillation frequency with the incoherent synchrotron frequency, reduced by space charge, a threshold space charge parameter $\Sigma_{th}$ can be obtained, which depends only on the bunch length.   
In \cite{Boine-Frankenheim:2005fk} the simple criteria were applied to single and dual rf buckets and confirmed by particle tracking studies, using a conventional grid-based space charge solver. For the solver a cut-off harmonic $h_c$ was chosen well below the actual cut-off wavelength of typical beam pipes, in order to avoid artificial numerical heating. 

The effect of the cut-off frequency or the resonant frequency of a broadband impedance on the loss of Landau damping was studied in \cite{Karpov2021} using the Lebedev equation as well as particle tracking. The criterion of an emerging van Kampen mode was used to determine the
threshold for loss of Landau damping.
An analytic expression, which scales inversely
with the cutoff frequency, was obtained for the threshold in the case of dipole oscillations in a single rf bucket. This expression results in a threshold at zero intensity in the limit $h_c\rightarrow\infty$. 
The finite thresholds obtained in previous studies are
attributed to the artificially lowered cut-off frequency 
in the numerical schemes used.

In this study we first characterize a longitudinal particle tracking code using a conventional grid-based space charge solver, in the limit $h_c\rightarrow\infty$. In addition to common convergence studies (with $h_c$, for example) we also use conserved quantities of the Vlasov system to estimate the accuracy of the simulations. We introduce a grid-less spectral solver as an alternative simulation approach. Such spectral or symplectic schemes have already been employed in particle tracking with two-dimensional space charge \cite{Qiang2017b}. Here we will use such a scheme in the longitudinal plane only. The advantage of spectral schemes, in particular for the case $h_c\rightarrow\infty$, lies in the much better conservation of the total energy, entropy and phase space, as we will show. An attempt to interpret our findings from the simulations is given in \autoref{sec:interpretation}.   


The structure of the paper is as follows.
In \autoref{sec:Model} the underlying longitudinal beam physics model as well as the problem description are given. Analytical approximations for coherent and incoherent frequencies as well as the resulting simple estimates for the loss of Landau damping are briefly summarized in \autoref{sec:tunes}. Grid-less and grid-based solvers for longitudinal space charge are described in \autoref{sec:grid_less} and  \autoref{sec:grid_based}. In \autoref{sec:entropy} we discuss and quantify the non-conservation of the total energy, entropy, and bucket area in grid-based and grid-less schemes. The application of the simulation schemes to the loss of Landau damping is the subject of \autoref{sec:simulations}.  

\section{Longitudinal beam dynamics model}\label{sec:Model}

The beam dynamics model used in this study is represented by the reduced Hamiltonian
in the coordinates $z$ (distance) and $\delta$ (relative momentum deviation) 
\begin{equation}\label{eq:H} 
H=-\frac{1}{2}\eta_0\delta^2-\frac{q}{2\pi R \beta_0^2 E_0}\int_0^{z} V(z) dz
\end{equation}
for a bunch in a stationary rf bucket where the voltage profile $V(z)$ is determined by the rf voltage $V_{\rm rf}=V_0\sin(z/R)$ and the space charge voltage
\begin{equation}\label{eq:V} 
V(z)=V_{\rm rf}(z) + \frac{\beta_0c}{2\pi R}\sum_h Z_h \lambda_h e^{ihz/R}.
\end{equation}
$\eta_0=1/\gamma_t^2-1/\gamma_0^2$ is the zero order slip factor, $\delta$ is the momentum deviation, $q$ the particle charge, $R$ the ring radius, $\beta_0$ and $\gamma_0$ the relativistic parameters, $E_0=\gamma_0 m c^2$. For simplicity, we assume a rf harmonic equal to one, corresponding to just one bunch in the ring. This does not restrict the application of our paper, as we will deal with space charge only and provide all main results in the form of dimensionless parameters. 

The space charge impedance can be approximated by \cite{Al-khateeb:2001yq}   
\begin{equation}\label{eq:sc_imp}
Z^{sc}(f)=-i h \frac{Z_0g}{2\beta_0\gamma_0^2}\frac{1}{1+(h/h_c)^2} 
\end{equation}
where $h=f/f_0$ is the harmonic number and $f_0$ is the revolution frequency, $Z_0=377\;\Omega$, $g=1+2\ln(b/a)$ ($b$: pipe radius, $a$: beam radius). 
The cut-off harmonic $h_c\approx \gamma_0 R/(b \beta_0)$ in the above expression is a fit parameter (see also \cite{Venturini2008}). For coherent bunch modes in hadron synchrotrons at injection energies, scale lengths of the order of the cut-off harmonic (typically $h_c\approx 1000$) usually play no role and the space charge induced voltage can be approximated as 
\begin{equation}
V_s(z)=-q\beta_0cRX_s\frac{\partial\lambda}{\partial z}
\end{equation}
where $X_s=g/(2\epsilon_0\beta_0 c\gamma_0^2)$ is the space charge reactance.
A space charge parameter is usually defined as \cite{Boine-Frankenheim:2005fk}
\begin{equation}
    \Sigma=\frac{1}{V_0/V_{s0}-1}
\end{equation}
where $V_0$ is the rf amplitude and $V_{s0}$ the space charge induced voltage amplitude.   

For a local elliptic bunch distribution 
\begin{equation}
    f(H)=f_0 \sqrt{H_m-H}
\end{equation}
where $H_m=H(0,z_m)$ and $z_m=l_b/2$ is the bunch half-length, 
this leads to the following expression 
\begin{equation}
    \Sigma=\frac{q^2\lambda_0}{\pi\gamma_0 m \beta_0 c \eta_0 \delta^2_m} X_s
\end{equation}
where the line density at the bunch center is $\lambda_0=3N_b/(2 l_b)$ and $\delta_m$ is the (half) momentum spread. The local elliptic or Hofmann-Pedersen bunch distribution can be analytically matched with space charge in arbitrary rf buckets. 

\section{Incoherent and coherent dipole frequencies}\label{sec:tunes}

For the case of the local elliptic bunch distribution
we briefly summarize some of the main analytic expressions, following the approach outlined in \cite{Hofmann1979a} and used for example in \cite{Boine-Frankenheim:2005fk}.  

In the presence of space charge and for small synchrotron oscillation amplitudes  $\hat{\phi}=\hat{z}/R$ the synchrotron frequency is
\begin{equation}\label{eq:syn_spread}
\frac{\omega_s}{\omega_{s0}}\approx\sqrt{\frac{1}{1+\Sigma}}\left(1-\frac{\hat{\phi}^2}{16}\right)
\end{equation}
where the synchrotron frequency for $\Sigma=0$ and $\hat{\phi}=0$ is 
\begin{equation}\label{eq:syn_freq}
\omega_{s0}^2=\frac{q\eta_0 V_0}{2\pi m \gamma_0 R^2}.    
\end{equation}
The rigid dipole frequency can be approximated as
\begin{equation}\label{eq:dipole}
\frac{\Omega_0}{\omega_{s0}}\approx\left(1-\frac{\phi_m^2}{10}\right)^{1/2}
\end{equation}
where $\phi_m=z_b/R$ is the bunch half-length.

Landau damping of the dipole mode is usually assumed, if $\Omega_0$ lies within the band of incoherent synchrotron frequencies  
\begin{equation}
\omega_{s,\min}<\Omega_0<\omega_{s,\max}
\end{equation}
where $\omega_{s,\min}$=$\omega_s(\phi_m)$, $\omega_{s,\max}$=$\omega_s(0)$.
For the rigid dipole modes this leads to the threshold space charge parameter
\begin{equation}
\Sigma_{th}\approx\frac{\phi_m^2}{10}.    
\end{equation}
In hadron synchrotrons at injection energies the space charge parameters typically are above this threshold value and active damping of the bunch center oscillations is required and performed routinely. However, it is still important to understand the underlying Landau damping mechanism, also for higher-order bunch modes, where active damping is more difficult.     

For Gaussian bunch distributions one has to keep in mind that for amplitudes close to the rms bunch length $\sigma$ and for a sufficiently large space charge parameter the synchrotron frequency is higher than the one for particles at the bunch center
$\omega_s(\sigma)>\omega_s(0)$ \cite{Boine-Frankenheim:2007fk}. Therefore, the criteria cannot be applied straightforwardly to Gaussian bunches. In \cite{Boine-Frankenheim:2007fk} it was shown that the loss of Landau damping in Gaussian bunches occurs at much lower space charge parameters, compared to elliptic bunches. Numerical errors in macro-particle simulation codes can drive an initial bunch distribution towards an 'equilibrium' Gaussian bunch distribution, which then shows a loss of Landau damping. Therefore, the effect of 'artificial collisions' in such simulations has to be kept at sufficiently low levels.          

\section{Grid-less multi-particle tracking with space charge}\label{sec:grid_less}

The goal of our numerical integration scheme
is to approximate the solutions of the Vlasov equation, corresponding to the Hamiltonian \autoref{eq:H}, for the bunch distribution $f(z,\delta,t)$ as closely as possible.  
For the approximation of the continuous distribution function
we consider a macro-particle distribution represented by shape functions in position space and delta functions in momentum space
\begin{equation}
f(z,\delta,t)=\frac{Q}{q\Delta}\sum_j^{N_p}S(z-z_j)\delta(\delta-\delta_j) 
\end{equation}
where $N_p$ is the number of macro-particles, $\Delta$ is the extension of a macro-particle in position space and $Q=qN_b/N_p$ is the charge of a macro-particle.   
The longitudinal multi-particle Hamiltonian, including possible higher-order contributions to the slip-factor, is
\begin{equation}\label{eq:Hmult}
H=-\sum_j^{N_p} \left(\frac{1}{2}\eta_0\delta_j^2 + \frac{1}{3}\eta_1\delta_j^3+\dots\right)-\frac{q}{2\pi R \beta_0^2 E_0}\sum_j\int_0^{z_j} V(z_j) dz_j
\end{equation}
with the rf and space charge voltages 
\begin{equation}
V(z_j)=V_{\rm rf}(z_j) + V_{\rm sc}(z_j).
\end{equation}
The smoothed space charge voltage acting on particle $j$ is obtained from 
\begin{equation}
V_{\rm sc}(z_j) = \sum\limits_{h=-N_h}^{N_h} Z_h \lambda_h S(h) e^{i h z_j / R}
\end{equation}
where $Z_h$ is the space charge impedance at harmonic $h$ and we use $N_h$ harmonics. $S(h)$ is the Fourier transform
of tent-like shape functions $S(z)$ in position space
\begin{equation}
S(h)=\sinc^2(h \Delta / 2).
\end{equation}
The line density harmonics are obtained from the macro-particle positions $z_l$ as
\begin{equation}
\lambda_h = Q S(h)\sum\limits_{l=0}^{N_p} {{e^{ - ihz_l/R }}}.
\end{equation}
This grid-less approach is similar to the one used in spectral PIC codes (see \cite{Decyk2011b}). 

For a given voltage function  
the longitudinal equations of motion for particle $j$ are
\begin{equation}
z'_j=-\eta(\delta_j) \delta_j, \quad \delta'_j=\frac{q}{2\pi R \beta_0^2 E_0} V(z_j)  
\end{equation}
where the slip factor is 
\begin{equation}
\eta(\delta)=\eta_0+\eta_1 \delta + \eta_2 \delta^2 + \ldots     
\end{equation}
The symplectic difference or mapping equations for one turn are
\begin{eqnarray}
\delta_j^{n+1}=\delta_j^{n}+\frac{q}{\beta_0^2 E_0} V(z_j^n) \\ 
z_j^{n+1}=z_j^{n}-2\pi R \eta(\delta_j^{n+1}) \delta_j^{n+1}. 
\end{eqnarray}
For a multi-particle space charge voltage, with dependencies on the full vector $\bf z$ of particle positions,    
the time advance remains symplectic if the interaction matrix $L_{i,j}=\frac{\partial H({\bf z})}{\partial z_i\partial z_j}$ is symmetric (see \cite{Qiang2017b}, for example).
\begin{equation}
L_{i,j}\propto\frac{\partial V(z_i)}{\partial z_j} =  i\frac{q}{R} \sum_{h=0}^{N_h} h S^2(h)\left(Z_h^* e^{-ih/R(z_i-z_j)}-Z_he^{ih/R(z_i-z_j)}\right)              
\end{equation}
For purely imaginary impedances, like for the space charge impedance,  this is the case for the presented grid-less approach, but not for grid-based PIC.

\section{Grid-based approach}\label{sec:grid_based}

In grid-based approaches, nonphysical grid forces can arise from aliasing, which occurs when the continuous particle density has spatial variations less than the grid spacing.  Such variations cannot be resolved, but they get mapped onto longer wavelengths \cite{Decyk2011b}. 

In plasma simulations, grid-induced artificial heating can be avoided by choosing a sufficiently small grid spacing to resolve the Debye length. In longitudinal beam simulations with the space-charge impedance \autoref{eq:sc_imp}, a similar role is played by the cut-off wavelength $2\pi R/h_c$.      

We use a grid with an equidistant spacing $\Delta z=L/N_z$, where $N_z$ is the number of grid points and $L=2\pi R$ is the length of the model accelerator. 
The line charge density at grid point $z_u=u \Delta z$ and time step $n$ is 
\begin{equation}
\lambda(z_u)=Q\sum_j^{N_p} S(z_u-z_j)   
\end{equation}
where $S(z)$ is the particle shape (tent-like), $z_j$ the particle position, $N_p$ is the number of macro-particles and $Q=q N/N_p$ is again the macro-particle charge ($N$ is the number of beam particles).  

The complex voltage amplitude $V$ induced 
by a modulation of the line density $\lambda$ at harmonic $h$ is   
\begin{equation}
V_h=Z_h^{sc} I_h
\end{equation}
where $I_h=\beta_0 c \lambda_h$ is the current amplitude and
\begin{equation}
\lambda_h=\sum_u \lambda(z_u) \exp(i h z_u / R).
\end{equation}
After Fourier back-transforming, the voltage acting on a particle $j$ is then obtained via 
\begin{equation}
V(z_j) =\sum_u S(z_u-z_j) V(z_u).
\end{equation}
The grid-based approach, employing FFTs, requires fewer operations compared to the grid-less approach. Therefore, it is preferred for time-consuming parameter scans. However, one must ensure $\Delta z \ll L/h_c$ to avoid grid effects. In order to limit the number of grid points $N_z$ a reduced $h_c$ has to be used, and the limit $h_c\rightarrow 0$ cannot be resolved.   

\section{Numerical energy, entropy and phase space non-conservation}\label{sec:entropy}

The Hamiltonian of the multi-macro-particle system \autoref{eq:Hmult}
with space charge represents the conserved total energy (below transition energy). 
Any change in total energy is due to numerical integration errors. 
This holds as long as the rf voltage is treated as time-independent, as it is the case in our model. From the symplectic time-advance we can expect a bounded energy error.   
Grid-based PIC codes usually conserve momentum but not energy.

In \cite{Brackbill2016} the energy error is found to scale inversely proportional to the number of particles per cell, which in our case is
\begin{equation}
N_c=\frac{N_p}{h_c}
\end{equation}
the number of particles per cut-off wave length and the energy error can be expected to scale with $N_c$ as
\begin{equation}
\Delta W\propto N_c^{-1}.     
\end{equation}

In grid-based schemes one usually ensures that particles move less than one grid cell per time step (Courant-Friedrichs-Lewy condition). In \cite{Brackbill2016} the energy error related to the time step was shown to scale with the number of cells an average velocity particles travels per time step, which in our case is  
\begin{equation}\label{eq:C}
C=v\frac{\Delta t}{\Delta z}\approx (1+\Sigma)\eta_0\delta_m N_z N_T    
\end{equation}
where we used $v=(1+\Sigma)\eta_0\delta_m$ for the characteristic velocity, including space charge. $N_T$ is the number of turns per time step. For small energy error $N_c\gg 1$ and $C<1$ should be chosen. The relative energy error $\Delta W/W_0$ (initial total energy $W_0$) for a given total simulation time should always remain well below 0.01, which can lead to very large macro-particle numbers and/or small time steps.  

\begin{figure}[htb]
    \centering
    \includegraphics[width=\linewidth]{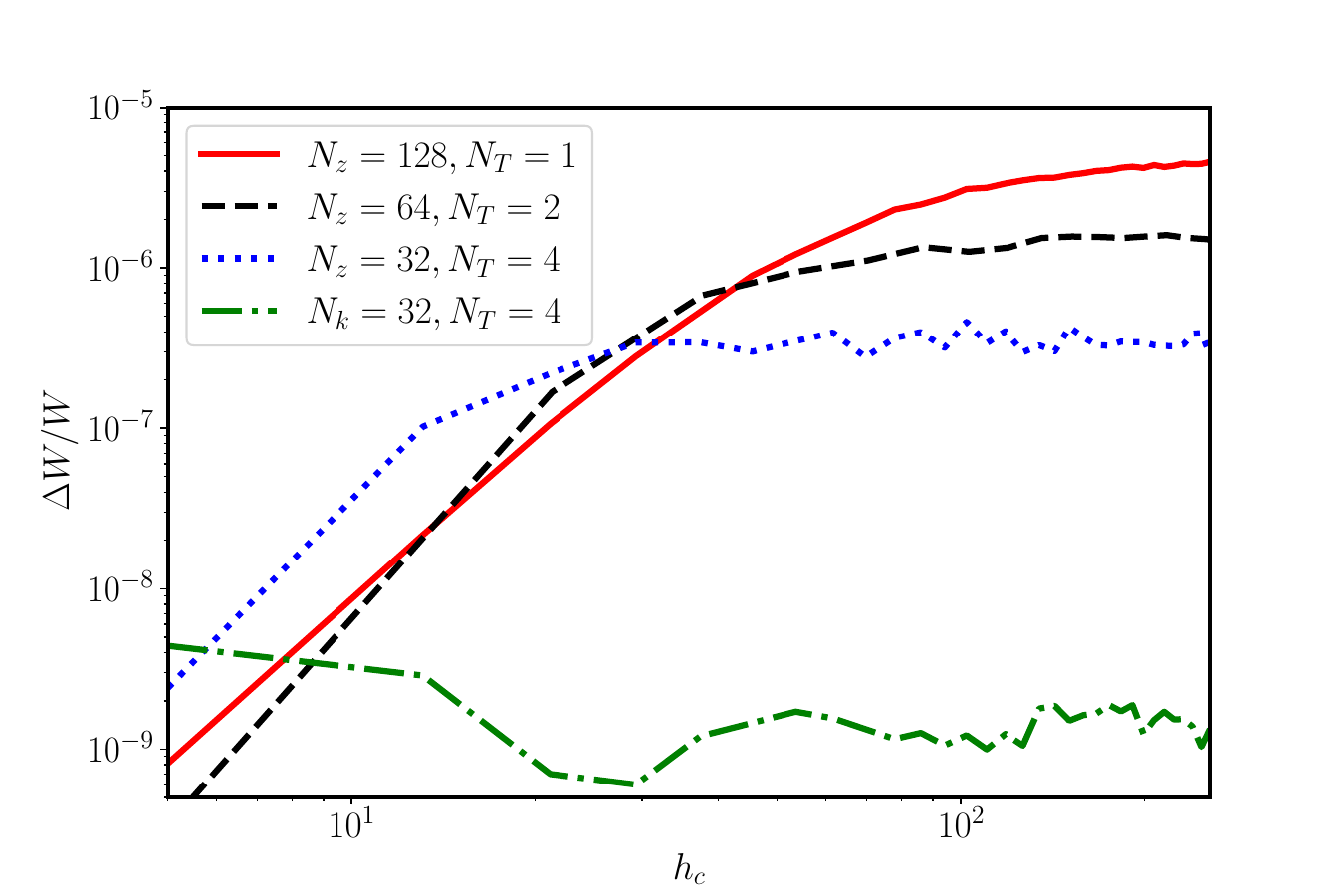}
    \caption{Relative energy growth per turn as a function of the cut-off harmonic $h_c$.
    Shown are three results from grid-based simulations for different $N_z$. In addition an example result for the spectral scheme ($N_k=32$) is shown. The space charge parameter is $\Sigma=1$, the macro-particle number $N_p=100000$ and the time step $C\approx 1$.} 
    \label{fig:scan_hc}
\end{figure}

As an example case, we use a single bunch with an elliptical distribution, initially matched with $\phi_m=60^o$ in a rf wave for a space charge parameter $\Sigma=1$.  
In \autoref{fig:scan_hc} we plot the relative energy growth for one turn obtained from different simulation runs as a function of the chosen cut-off harmonic $h_c$. 
If one considers that a full simulation run over many synchrotron oscillation periods typically requires $10^6$ turns, this results in a tolerable relative energy growth per turn of about $10^{-8}$. 

The grid-based schemes show the expected $1/N_c$ increase up to $N_z\approx h_c$. For $N_z<h_c$ the energy growth saturates, as effects above the cut-off wavelength cannot be resolved anymore. Although $C\approx 1$ is the same for all simulations and we would expect a similar energy growth (for $N_z<h_c$), the simulation with the lowest grid resolution $N_z=32$ shows a larger energy growth. The energy growth obtained from the spectral simulation scheme for mode numbers $N_k=32$ (all other parameters are the same) is negligible small for all $h_c$.   
This statement also hold for larger $N_k$ (not shown).  

The energy growth is the most precise indicator of numerical errors, in the case where energy is a conserved quantity, like for our model Hamiltonian. In more general cases we can use the fact that the Vlasov equation conserves phase space as well as entropy defined as 
\begin{equation}\label{eq:entropy}
S=\int f \ln f  dzd\delta 
\end{equation}
for the distribution $f(z,\delta,t)$. However, due to the finite grid spacing and/or finite number of macro-particles fine-grained information will be lost eventually and numerical integrators for the Vlasov equation do not conserve $S$ (see for example \cite{Manfredi1997}). 
For PIC codes the entropy can be obtained directly via \autoref{eq:entropy} by interpolating the macro-particle distribution onto a 2D grid in $(z,\delta)$ space (see \cite{Liang2019}).  
For grid-less multi-particle schemes as well as for conventional PIC, a more appropriate approach is the entropy estimator 
\begin{equation}
S_E=\frac{d}{N_p}\sum_{j=1}^{N_p}\ln(R_j)+\ln\left(\frac{\pi^{d/2}}{\Gamma(d/2+1)}\right)+\ln(N_p)+\gamma  
\end{equation}
where $R_j$ is the euclidian distance to the nearest neighbor of the $j$th particle in $(z,\delta)$ phase space \cite{Mirsa2003}. This is the approach that we will use in the following.

First, it is useful to recall some of the analytical estimates for the entropy and its evolution.
For a matched bunch the entropy $S_m$ is related to the bunch phase space area $A_m$ through \cite{Lawson1973}
\begin{equation}
\frac{S_m}{N}=\ln(A_m)+\textrm{const.}  
\end{equation}
where the matched bunch area is $A_m=\pi z_m \delta_m$ and $N$ is the number of particles.
In \cite{Struckmeier1996} this relationship was generalized to the evolution of beam distributions, which remain rms matched  
\begin{equation}
    \frac{1}{N}\frac{dS_m}{dt}=\frac{d}{dt}\ln A_m
\end{equation}
where we assume that the space charge contribution can be neglected. For short time intervals or small changes we obtain
\begin{equation}\label{eq:entropy_vs_bunch}
\Delta S_m \approx \frac{\Delta A_m}{A_m}.     
\end{equation}

\begin{figure}[htb]
    \centering
    \includegraphics[width=\linewidth]{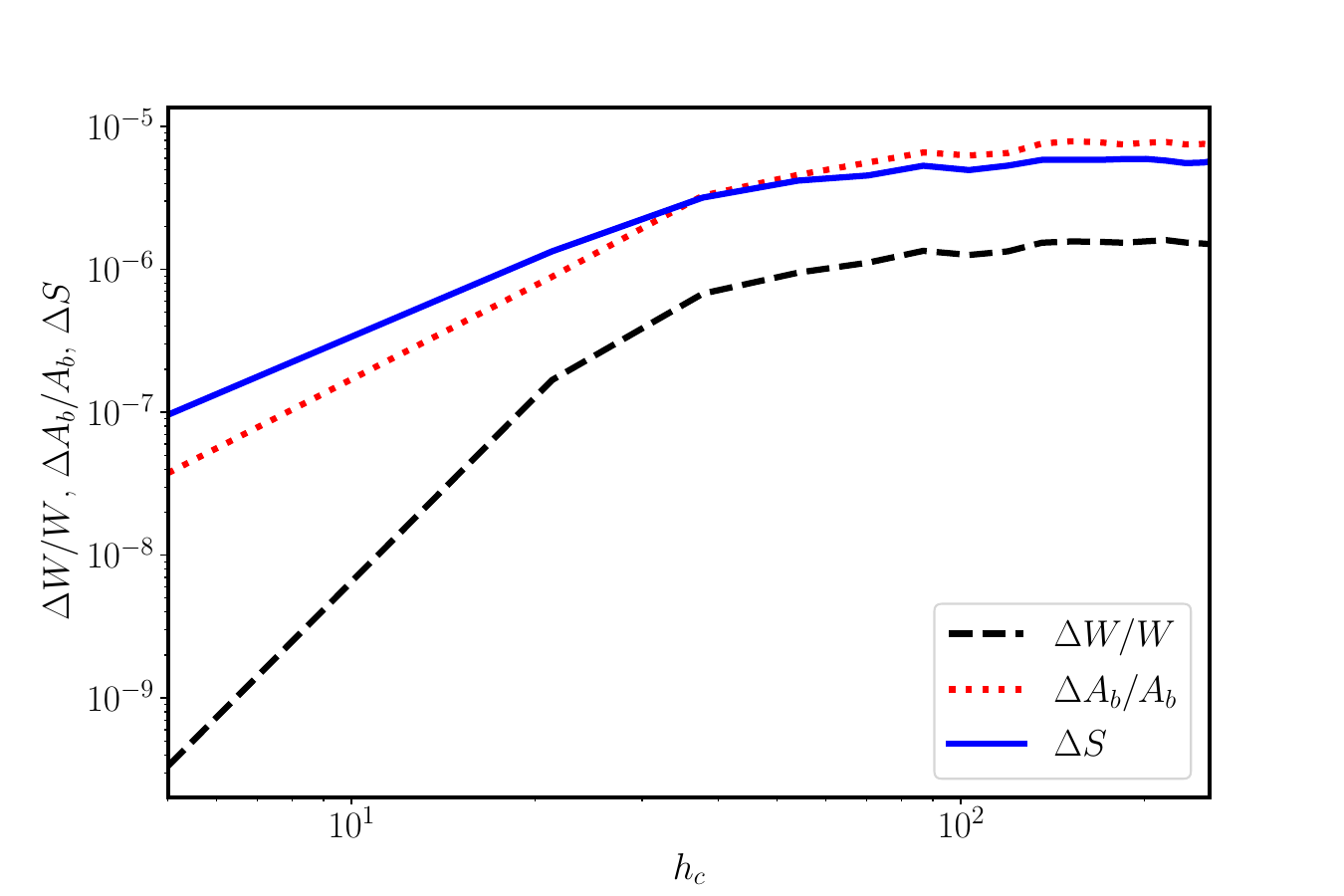}
    \caption{Relative bunch area and entropy change per turn as a function of the cut-off parameter $h_c$ for a matched bunch with $\Sigma=1$ from grid-based simulation with $N_p=100000$ macro-particles and $C\approx 1$ ($N_z=64, N_T=2$). For comparison, the relative energy growth, given already in \autoref{fig:scan_hc}, is also shown.}\label{fig:scan_hc_2}
\end{figure}

\autoref{fig:scan_hc_2} shows the relative change in the bunch area per turn together with the entropy change per turn from a grid-based simulation ($N_z=64$) of a matched bunch ($\Sigma=1$) as a function of the cut-off parameter $h_c$. For comparison we show the relative energy growth, given already in \autoref{fig:scan_hc}. For the example parameters, similar to the relative energy, we observe an increase with $h_c$ and a saturation. Furthermore, the relative bunch area growth and the entropy growth are qualitatively similar, as predicted by \autoref{eq:entropy_vs_bunch}.   

\begin{figure}[htb]
    \centering
    \includegraphics[width=\linewidth]{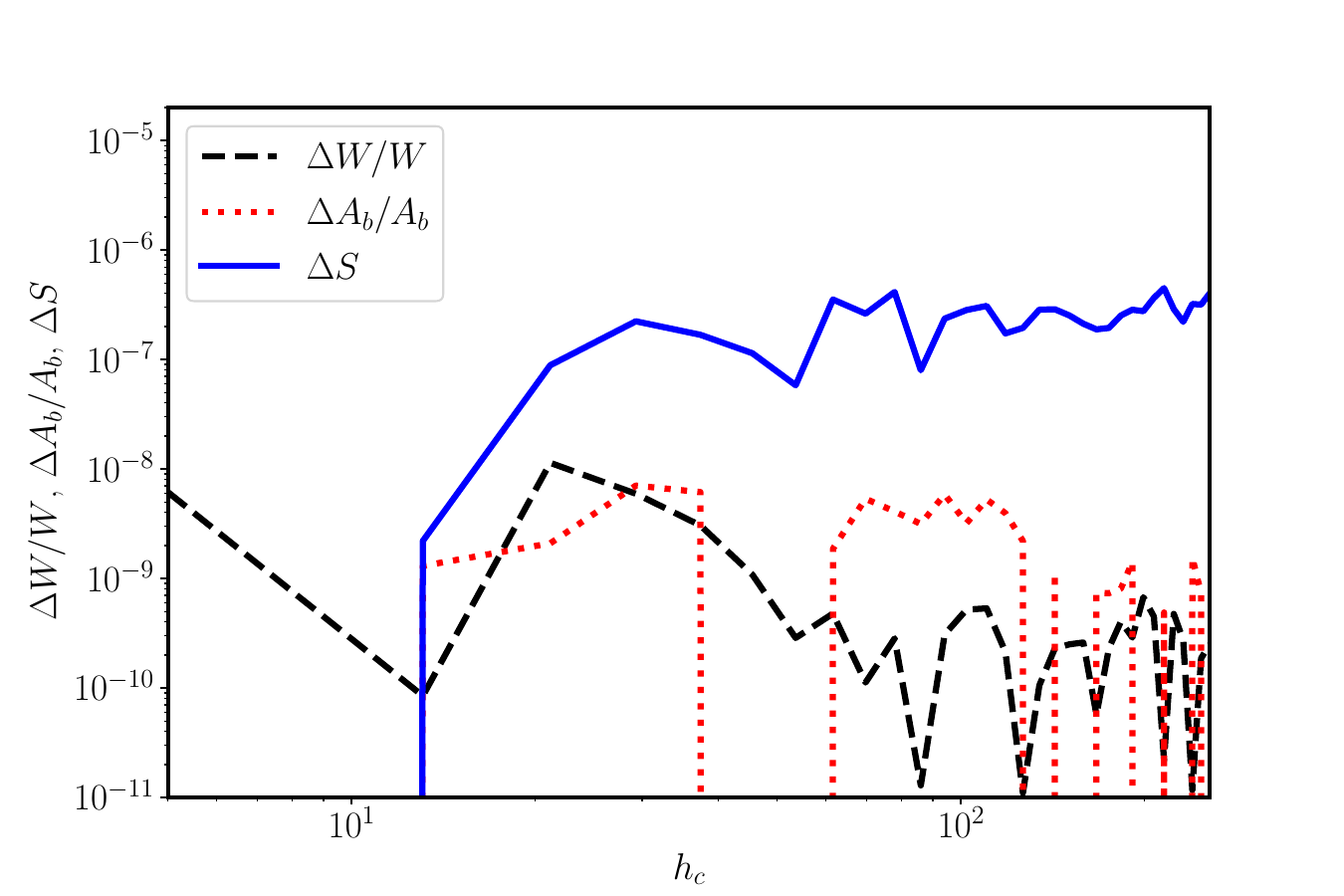}
    \caption{Energy, bunch area and entropy change per turn as a function of $h_c$ for a matched bunch ($\Sigma=1$) from a spectral simulation with $N_p=100000$ macro-particles and $C\approx 1$ ($N_k=32, N_T=4$).}
    \label{fig:scan_hc_spec}
\end{figure}

For the spectral simulation scheme (\autoref{fig:scan_hc_spec}) we find a more or less constant bunch area and energy (below the noise level). There is a noticeable entropy growth for $h_c>N_k$, still well below the level of the grid-based scheme.         

\begin{figure}[htb]
    \centering
    \includegraphics[width=\linewidth]{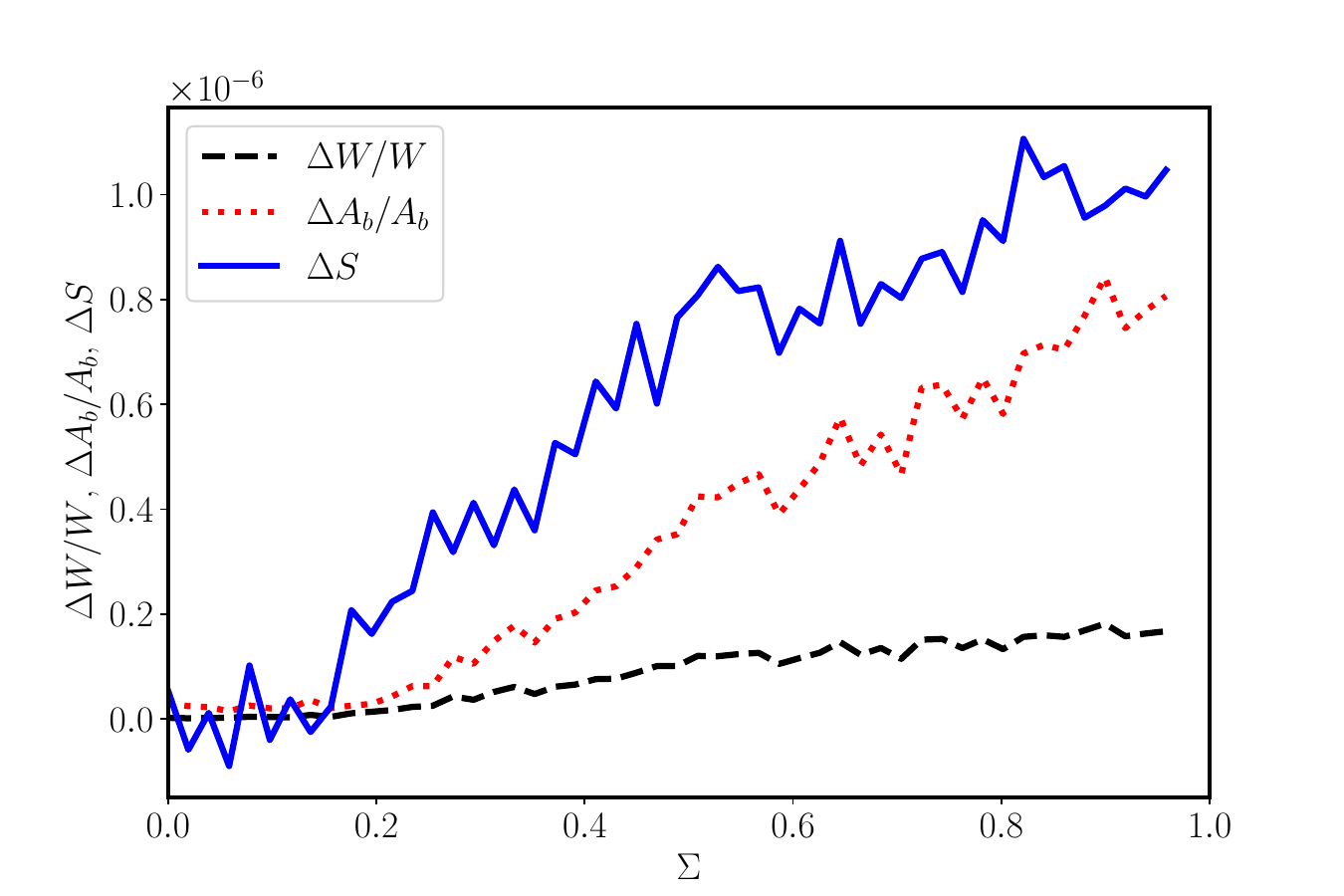}
    \caption{Energy, bunch area and entropy change per turn as a function of the space charge parameter $\Sigma$ and fixed $h_c=20$ for a matched bunch from grid-based simulation with $N_p=100000$ macro-particles and $C\approx 1$ ($N_z=64, N_T=2$).}
    \label{fig:scan_sigma}
\end{figure}

\begin{figure}[htb]
    \centering
    \includegraphics[width=\linewidth]{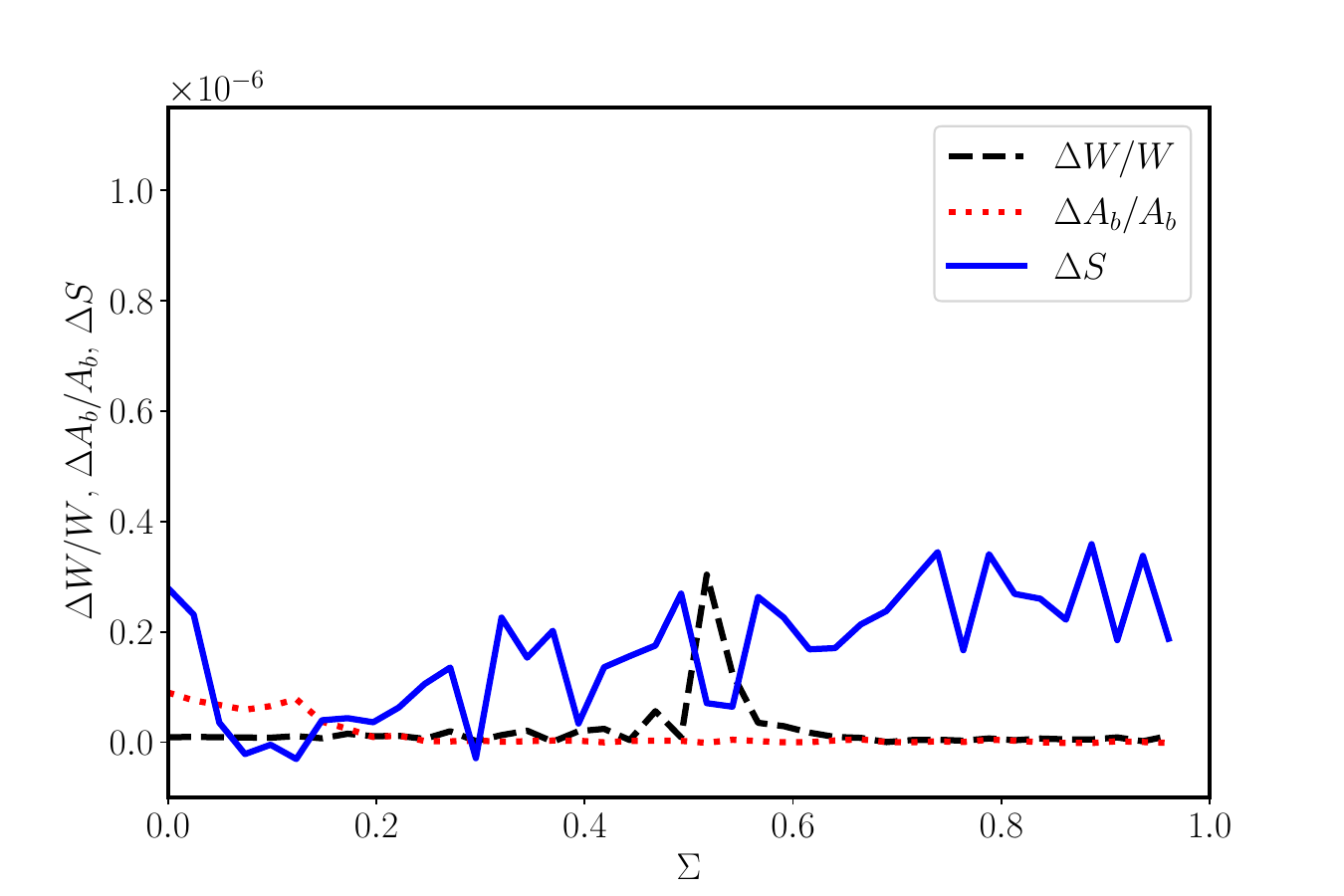}
    \caption{Energy, bunch area and entropy change per turn as a function of the space charge parameter $\Sigma$ and a fixed $h_c=20$ for a matched bunch from a spectral simulation with $N_p=100000$ macro-particles and $C\approx 1$ ($N_k=32, N_T=4$).}
    \label{fig:scan_sigma_spec}
\end{figure}

Next, we will discuss the effect of $\Sigma$, keeping $h_c$ constant. 
\autoref{fig:scan_sigma} shows the relative energy and bunch area change per turn together with the entropy change per turn from a grid-based simulation ($N_z=64$) of a matched bunch as a function of the space charge parameter $\Sigma$ for fixed $h_c=20$. For the spectral scheme ($N_k=32$) the corresponding plot is shown in \autoref{fig:scan_sigma_spec}. Starting around $\Sigma\approx 0.2$, the grid-based simulation shows a linear growth of the energy with $\Sigma$, which is consistent $\Delta W\propto C$, as obtained also by \cite{Brackbill2016}. The entropy growth reacts first and remains about a factor of 2 above the relative bunch area growth. For the spectral scheme we observe a very good energy conservation (a still unexplained bump around $\Sigma\approx 0.5$) as well as bunch area conservation. A slight entropy increase is observed, well below the values for the grid-based scheme.

We can conclude that for our application in this study the most precise indicator for artificial numerical effects is the total energy growth.
In our specific case of longitudinal dynamics with space charge (no resistive impedance components or time-dependent rf fields), the total energy is the preferred validation quantity. 
It is checked for a given simulation duration in addition to the usual convergence checks.
The grid-based scheme shows a strong numerical growth of the total energy $\Delta W/W\propto h_c$ as well as $\propto\Sigma$.  
For the spectral scheme the growth of the total energy as well as of the bunch area is within the noise level. Especially for numerical studies of space charge effects for large $h_c$ or even in the limit of $h_c\rightarrow\infty$ spectral schemes should be preferred or used to validate results of grid-based simulations.  

Under more general conditions with resistive impedance components, for example, the total energy is not conserved, but the bunch area or entropy are remaining quality estimators. 
For matched bunches both quantities should be conserved and their growth is purely due to numerical errors, either grid-based or due to the finite macro-particle number. For the purpose of code validation the entropy should be preferred, as it is a strict invariant of the Vlasov equation.  For the chosen matched bunch example with space charge only, the increase of $\Delta S$ and the relative rms bunch area $A_b$ with $h_c$ and $\Sigma$ can be interpreted in a similar way.

\section{Simulation of Landau Damping with space charge}\label{sec:simulations}  

We use a spectral and a conventional (PIC) grid-based particle tracking code to integrate \autoref{eq:H}, \autoref{eq:V} and \autoref{eq:sc_imp}. 
The chosen machine and bunch parameters are close to the ones used in \cite{Boine-Frankenheim:2005fk}. A single bunch with an elliptical distribution, matched at $\phi_m=60^o$ to a rf wave is used as initial condition. The bunch intensity is varied between $\Sigma=0$ and $0.3$, similar to the conditions in the GSI SIS18 heavy-ion synchrotron at injection energy.           
The simulation time corresponds to a few 1000 synchrotron oscillations, and the time step is one turn, for a synchrotron tune of about 0.005, in our model synchrotron. 
For the grid-based solver we chose $N_z=256$ and $N_p/N_z\gtrsim 200$ macroparticles to resolve higher-order bunch modes and to limit the numerically induced increase of the bunch area to values well below a percent. The cut-off harmonic is chosen to be in the range $h_c=10-40$, well below $N_z$. 

$C$ is kept below $1$ and $N_c$ sufficiently high to ensure a numerical energy growth well below 0.01 for the grid-based scheme. This is possible because of the moderate space charge parameters. We can therefore use the faster grid-based simulation scheme and use the spectral scheme for validations only.

\begin{figure}[htb]
    \centering
    \includegraphics[width=\linewidth]{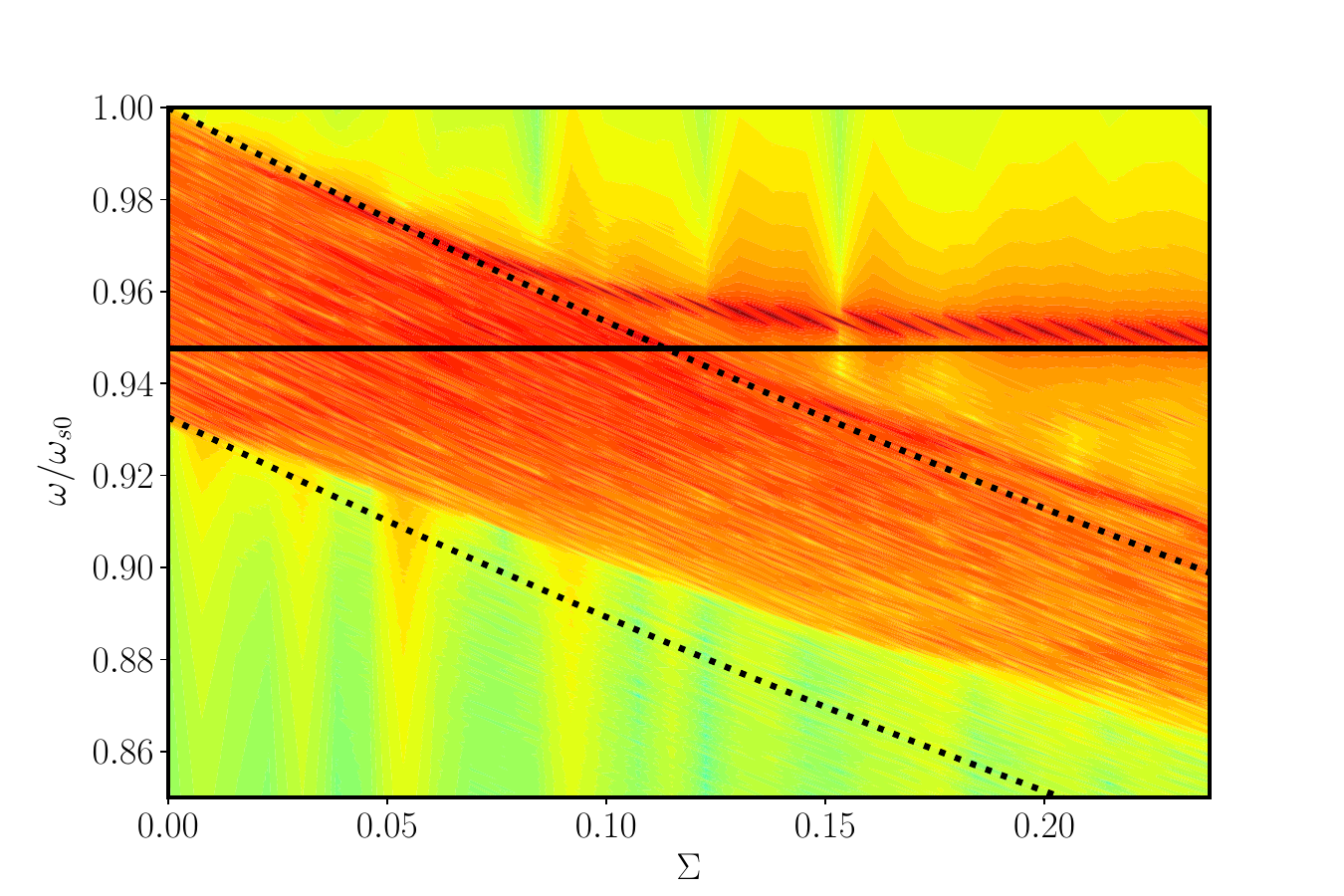}
    \caption{Bunch oscillation spectrum for different space charge parameters $\Sigma$ obtained from a grid-based simulation. The bunch length is kept at $\phi_m=60^0$. The cut-off harmonic is at $h_c=10$. The solid line is the dipole frequency. The upper and lower dotted curves represent the maximum and minimum synchrotron frequency in the bunch.}
    \label{fig:scan1}
\end{figure}
\begin{figure}[htb]
    \centering
    \includegraphics[width=\linewidth]{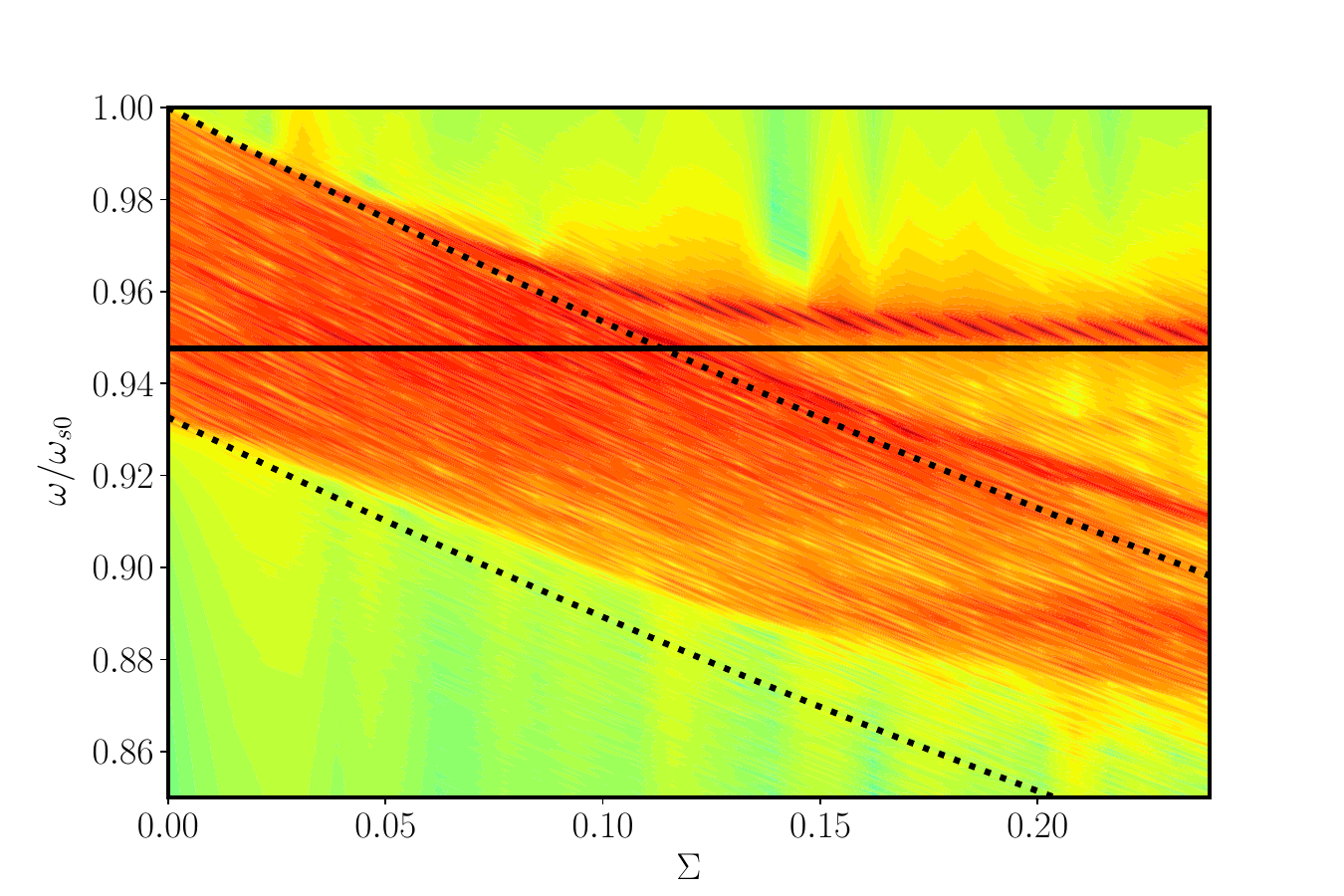}
    \caption{Bunch oscillation spectrum for different space charge parameters $\Sigma$ obtained from a grid-based simulation. The bunch length is kept at $\phi_m=60^0$. The cut-off harmonic is at $h_c=20$. The solid line is the dipole frequency. The upper and lower dotted curves represent the maximum and minimum synchrotron frequency in the bunch.}
    \label{fig:scan2}
\end{figure}

\autoref{fig:scan1} shows the result of a simulation scan with grid-based PIC for different space charge parameters $\Sigma$. Shown is
the obtained bunch center oscillation spectrum $\hat{Z}$ for different space charge parameters. The corresponding dipole frequency \autoref{eq:dipole} (solid line), synchrotron frequency \autoref{eq:syn_freq} (upper dotted curve) and synchrotron spread (lower dotted curve) \autoref{eq:syn_spread} are also shown. The cut-off harmonic is set to $h_c=10$. For $h_c=20$ the corresponding simulation results are shown in \autoref{fig:scan2}. Also, the result for $h_c=40$ (not shown) looks more or less identical. 
Our results indicate that simulation results obtained for a relatively low cut-off's, like the
$h_c=10-20$ used in the previous study (\cite{Boine-Frankenheim:2005fk}, Fig. 14), can predict the mode structure and Landau damping in the limit of $h_c\rightarrow\infty$. 

The scans also show that the rigid dipole frequency \autoref{eq:dipole} is a good approximation of the lowest-order coherent bunch oscillation mode for $\Sigma\gtrsim 0.2$. The coherent mode emerges from the incoherent frequency band at $\Sigma\approx 0.08$. The intersection of the rigid dipole frequency \autoref{eq:dipole} 
and the synchrotron frequency \autoref{eq:syn_freq}, used in \cite{Boine-Frankenheim:2005fk}, results in $\Sigma\approx 0.1$, which slightly overestimates the expected threshold for the loss of Landau damping from the scans ($\Sigma\approx 0.08$).   

\begin{figure}[htb]
    \centering
    \includegraphics[width=0.49\linewidth]{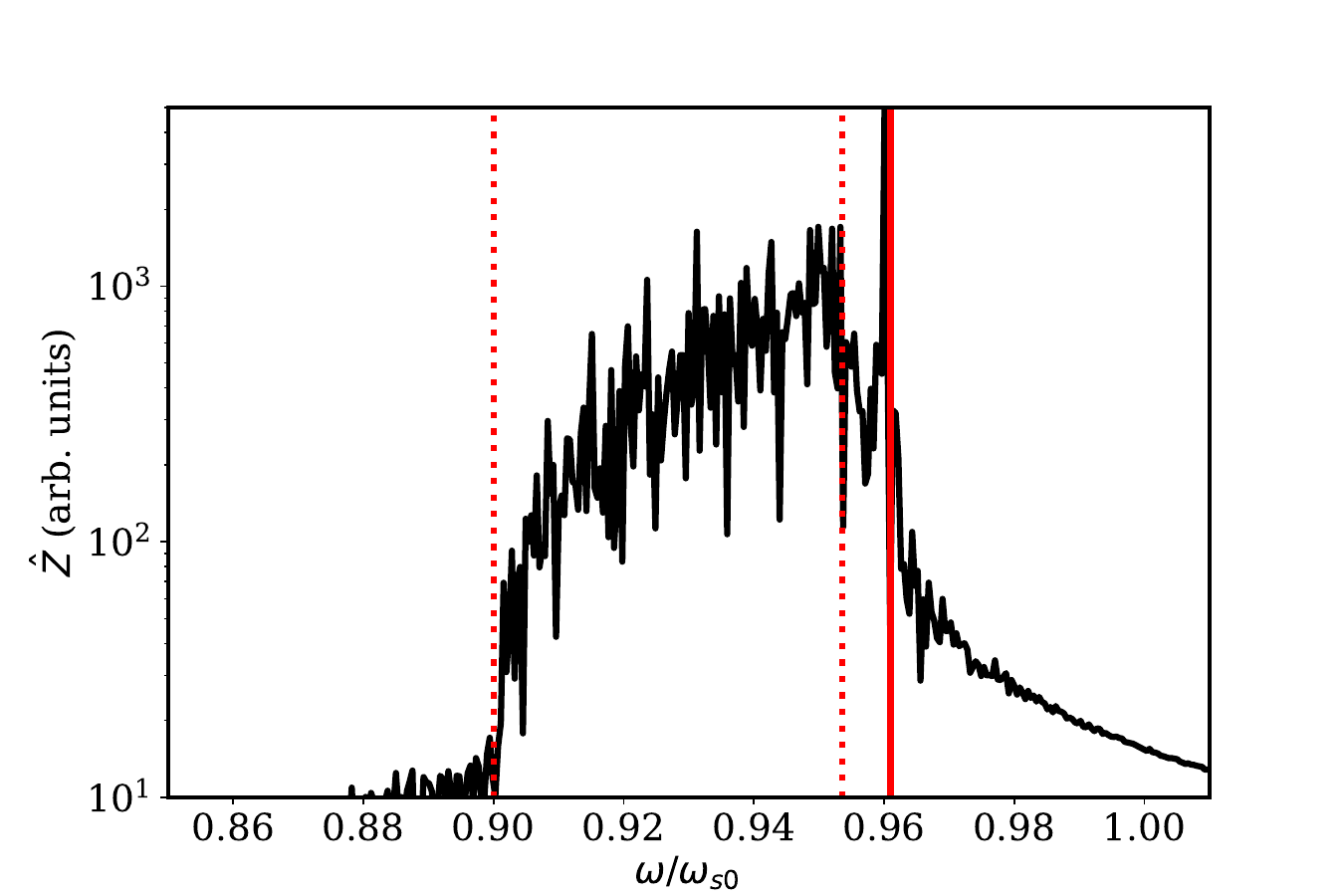}
     \includegraphics[width=0.49\linewidth]{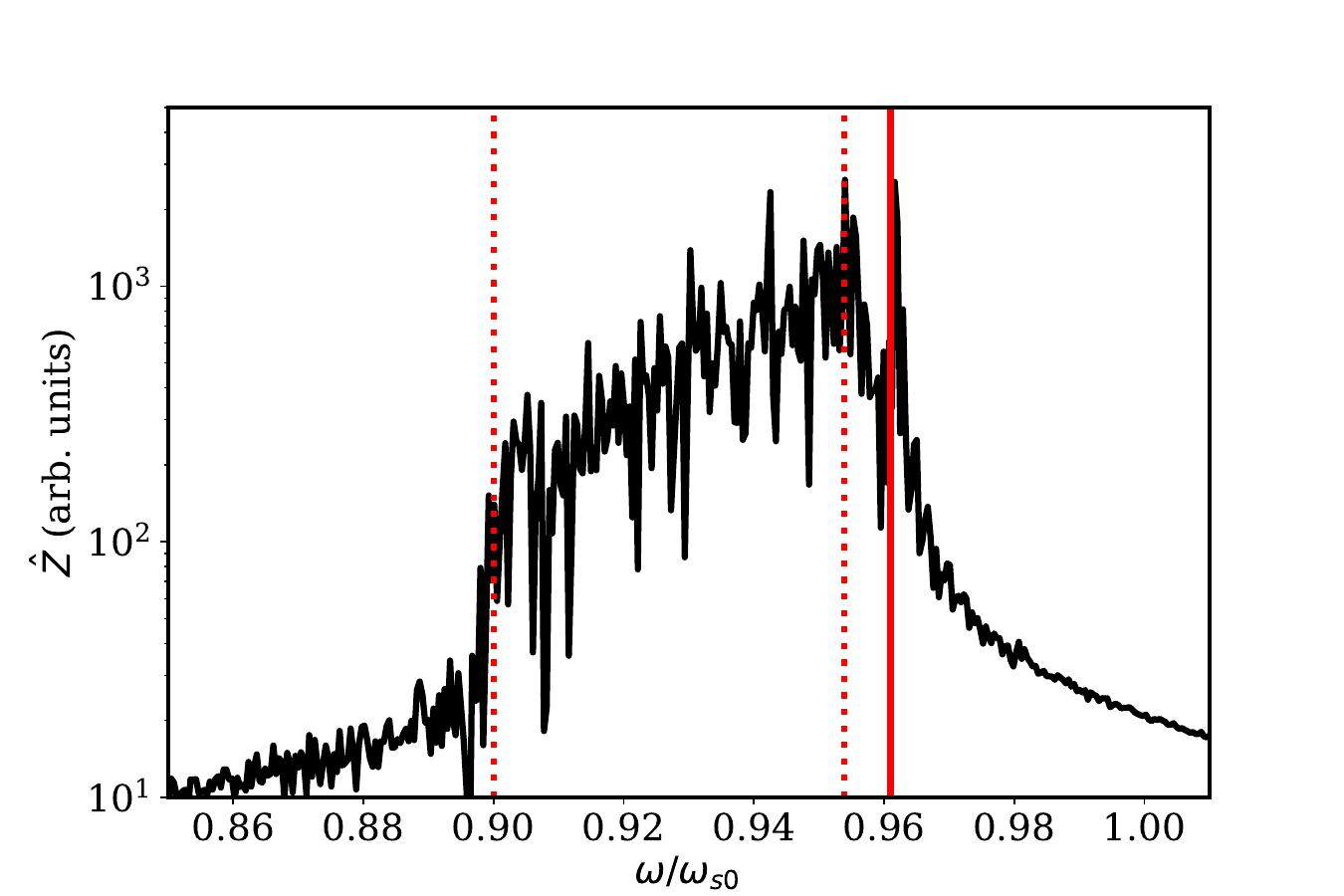}
    \caption{Bunch oscillation spectrum for $\Sigma=0.1$ obtained from a grid-based (left) and spectral (right) simulation. The bunch length is kept at $\phi_m=60^0$. The cut-off harmonic is at $h_c=20$. The solid red line indicates the dipole frequency. The two dotted lines indicate the maximum and minimum synchrotron frequency in the bunch.}
    \label{fig:spectra1}
\end{figure}

\begin{figure}[htb]
    \centering
    \includegraphics[width=0.49\linewidth]{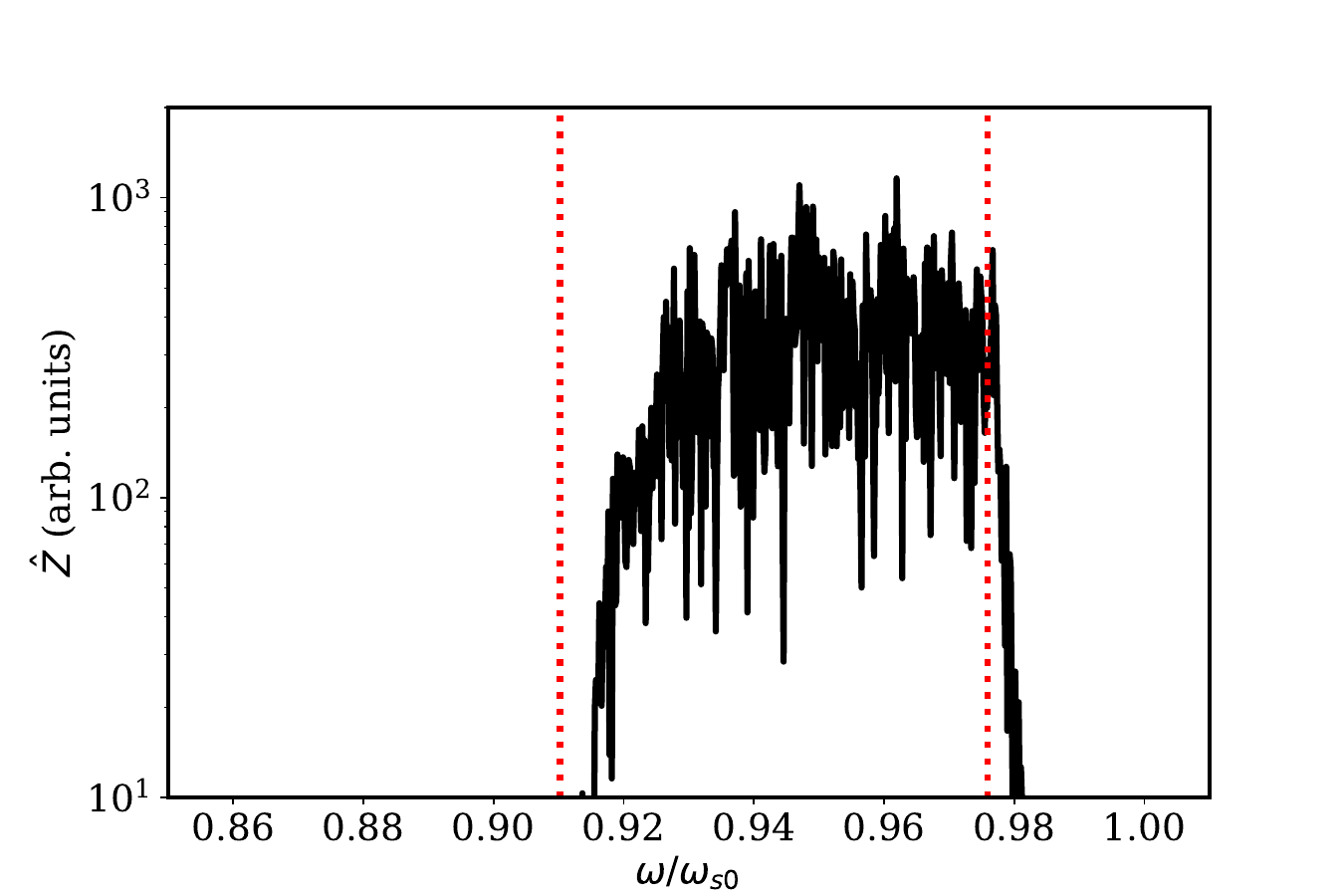}
     \includegraphics[width=0.49\linewidth]{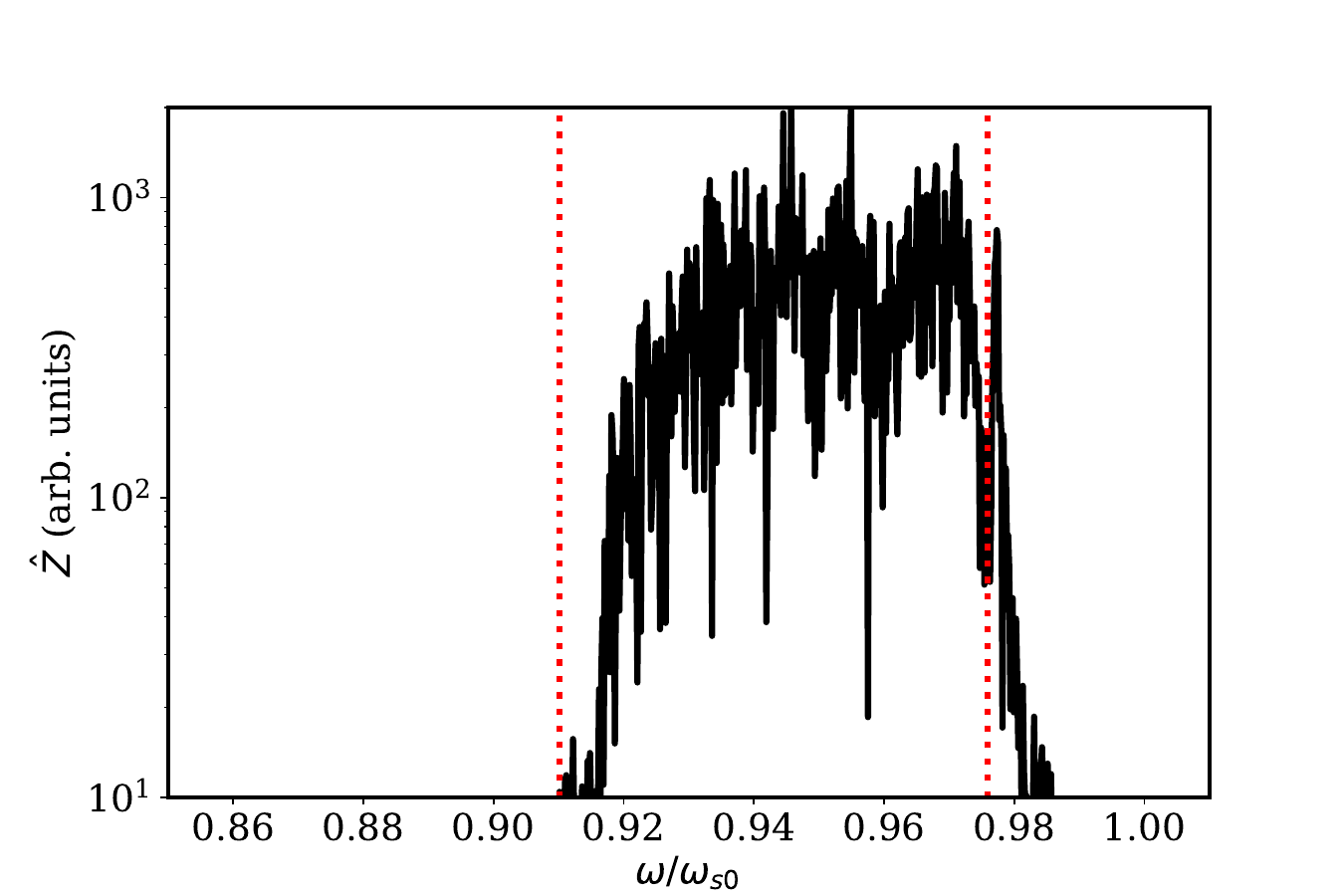}
    \caption{Bunch oscillation spectrum for $\Sigma=0.05$ obtained from a grid-based (left) and a spectral (right) simulation. The bunch length is kept at $\phi_m=60^0$. The cut-off harmonic is at $h_c=20$. The two dotted lines indicate the maximum and minimum synchrotron frequency in the bunch.}
    \label{fig:spectra2}
\end{figure}

In order to validate the spectra obtained from the grid-based scheme, we used the spectral scheme for two selected space charge parameters, $\Sigma=0.05$ and  $\Sigma=0.1$. They are located directly below and above the branching point at which the coherent mode emerges from the incoherent frequency band in the grid-based simulation scans shown previously. The results are shown in Figs. \autoref{fig:spectra1} and \autoref{fig:spectra2}. 
In the spectral scheme we used $N_k=128$ and a constant space charge impedance without cut-off.
For $\Sigma=0.1$ we can clearly observe the coherent lines at exactly the same positions.  For $\Sigma=0.05$ we observe the incoherent band only in both simulations (for the spectral scheme a low amplitude peak to the right of the incoherent band can be observed). 

\begin{figure}[htb]
    \centering
    \includegraphics[width=\linewidth]{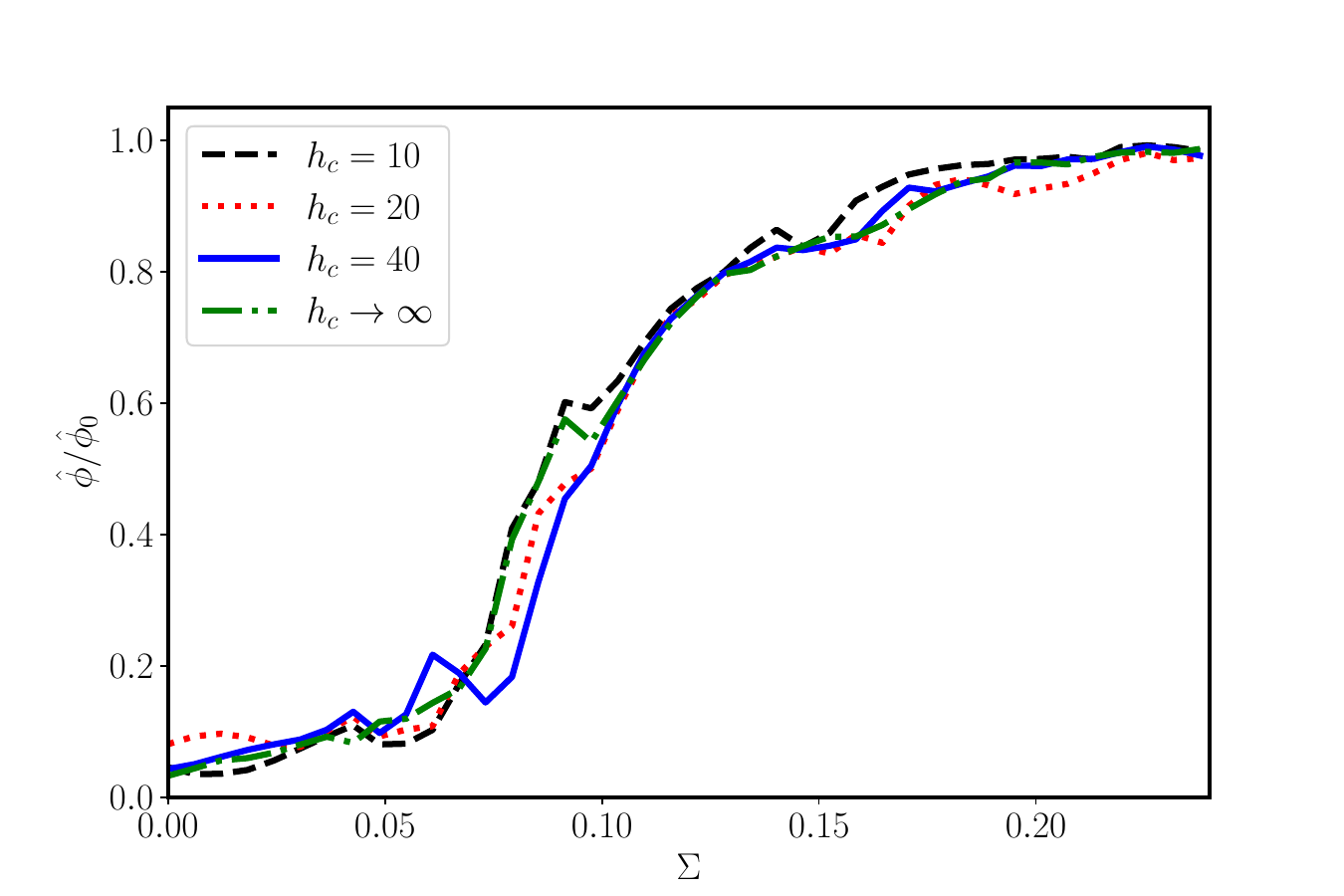}
    \caption{Residual oscillation amplitude for different space charge parameters $\Sigma$ and cut-offs $h_c$ obtained from grid-based and spectral simulation. The initial bunch length is $\phi_m=60^0$ and the offset $\hat{\phi}_0=1^0$.}
    \label{fig:scan3}
\end{figure}

In order to relate our results to the actual damping of a weak initial dipolar offset ($\hat{\phi}_0=1^0$) of a bunch (again with $\phi_m=60^0$) in a single rf bucket we performed simulations over short time intervals, just to resolve the damping of the initial offset and the remaining, residual oscillation amplitude. \autoref{fig:scan3} shows the residual amplitude $\hat{\phi}$ as a function of the space charge parameter from grid-based and spectral ('$h_c\rightarrow\infty$') simulations. We can clearly observe that the residual oscillation amplitude does not depend on $h_c$ and a "loss of damping" occurs at about $\Sigma\gtrsim 0.08$.  For $h_c<10$ the results start to differ, because the effective space charge parameter is reduced. The results in \autoref{fig:scan3} further confirm, that there is still an effective damping mechanism for $\Sigma < 0.08$, for the chosen example bunch parameters.      

\section{Interpretation of the simulation results}\label{sec:interpretation}

Our simulation study confirms the findings in \cite{Boine-Frankenheim:2005fk}. For the chosen example bunch, in the presence of space charge there is a small, but finite, threshold space charge parameter for the loss of Landau damping. The improved analysis, following the work in Ref. \cite{Karpov2021}, using the branching point of the coherent dipole mode in \autoref{fig:scan2},
indicates a threshold space charge parameter slightly below the one following from the simple criteria $\omega_s=\Omega_0$, with the rigid dipole frequency \autoref{eq:dipole}.   
Furthermore, the residual offset amplitudes shown in \autoref{fig:scan3} indicate a similar damping effect for increasing $h_c$.

Our findings are in contrast to the ones in Ref. \cite{Karpov2021} where analytically a vanishing threshold $\Sigma_{th}\propto 1/h_c$ is predicted (Eq. 53) for a purely inductive impedance (above transition). 
The findings are supported by semi-analytic solutions of the matrix equations for the linearized Vlasov equation in the coordinates of the unperturbed matched bunch.  
The prediction for $h_c\rightarrow\infty$ relies on the bunch beam transfer functions $G_{h}(\Omega)$, with the amplitude-dependent synchrotron frequency $\omega_s(\hat{\phi})$ as the source of Landau damping. 

We argue that the approximation used for $G_{h}(\Omega)$ in \cite{Karpov2021} is not valid for perturbations with wavelengths much smaller than the bunch length $l_b$ (or for $h\gg 2\pi R/l_b$, which in our simulation model results in $h\gg 3$). For large harmonics the $G_h(\Omega)$ used in \cite{Karpov2021} can be interpreted as the response of coasting beam-like modes to a (driven) low-frequency oscillation at $\Omega\approx\omega_{s0}$. Coasting beam modes are damped by the local momentum spread (see e.g. \cite{hofmann1995landau} or \cite{Boine-Frankenheim2000}, with space charge). For weak space charge the decay rate of a coasting-beam like line density modulation can be approximated as $\tau^{-1}_h\approx h f_0\eta_0\delta$, which is the inverse time needed by an average ion (with the local momentum spread $\delta$) to travel one wavelength. We can compare this decay rate with the synchrotron frequency and arrive at a maximum harmonic number  
\begin{equation}
h_{\max}\approx\frac{2 \pi \nu_s}{\eta_0\delta}.  
\end{equation}
For $h$ above this value the damping rate is faster than the dipole or synchrotron frequency and the framework used in \cite{Karpov2021} might not be applicable. For our simulation model 
$h_{\max}\approx 12$, which could explain the observed convergence of the results reached for $h=10$.  
In \cite{Karpov2021}
a further separation of the incoherent band and the coherent line with increasing $h_c$ is also reported in particle tracking simulations for a broadband resonator, centered at $h_c$ (\cite{Karpov2021}, Fig. 8). However, in our study, we focus on a purely imaginary space charge impedance with a smoothed cut-off at $h_c$.

\section{Conclusions}

For the longitudinal beam dynamics model (\autoref{eq:H} and \autoref{eq:sc_imp}) representing a single bunch affected by the space charge force below transition energy, we study the coherent lowest order dipole mode in a single rf bucket. The purpose of the study is to resolve the dependence of this mode and its Landau damping on the cut-off harmonic $h_c$ in the space charge impedance.     

Previous work (\cite{Hofmann1979a}, \cite{Boine-Frankenheim:2005fk}) indicated a low, but finite threshold space charge parameter $\Sigma_{th}\propto\phi_m^2$ (bunch half-length $\phi_m$) for an elliptic bunch distribution. Recently \cite{Karpov2021} a $\Sigma_{th}\propto 1/h_c$ dependence was found analytically and in particle tracking simulations for a broadband resonator impedance. Previous findings of finite thresholds were attributed to possible numerical issues or noise in tracking codes. 

In our study we point out that especially for the space charge impedance, grid-based solvers suffer from strong numerical energy growth $\Delta W \propto\Sigma h_c$ also called 'artificial heating'. A sufficiently large number of macro-particles has to be chosen to limit the growth for a given simulation time.   
In addition to the total energy, the entropy and the bunch area show similar growth, purely due to the numerical grid-particle coupling. We introduce the entropy as an alternative estimator for artificial effects, measuring the deviation from a solution of the Vlasov equation.     
As an alternative to conventional grid-baser solvers, we introduce a grid-less, spectral scheme for longitudinal beam dynamics simulations. This scheme shows a very good conservation of the total energy with $h_c$ and $\Sigma$, at the cost of a lower computational performance. We use this scheme for validation purposes.      

We use the simulation 
to confirm our previous study \cite{Boine-Frankenheim:2005fk}, in which we used a grid-based scheme with a cut-off $h_c\approx 10$ due to numerical reasons, as the actual cut-off for the real beam pipe in the synchrotron under investigation (GSI SIS18) would correspond to $h_c\approx 1000$. 
In our updated simulation study using grid-based and, for selected results, grid-free schemes, we follow the approach in \cite{Karpov2021}, to identify the threshold for the loss of Landau damping from scans of the bunch fluctuation spectrum. We identify the threshold space charge parameter with the branching point at which the coherent mode emerges from the incoherent frequency band. In addition, we also use the damping of an initial bunch offset as a function of the space charge parameter.

For our example bunch parameters we find that above $h_c\approx 10$ the results converge, leading to a finite threshold space charge parameter. The simple estimate $\Omega_0=\omega_s$ slightly overestimates the threshold parameter obtained. We argue, that there is a maximum contributing harmonic, close to the one found in our simulations, due to strong (faster than the synchrotron oscillation period) damping of coasting beam-like modes on the bunch.   

This is an important finding, as it demonstrates that we can resolve space charge effects in the limit of $h_c\rightarrow\infty$ using a low cut-off harmonic for numerical reasons, at least for low and moderate space charge parameters.    
For space charge dominated bunches and higher-order coherent modes or even microwave-type of instabilities, a grid-less macro-particle simulation scheme is absolutely required, as our studies indicate. Although we focus on longitudinal simulations, the same statement applies to the longitudinal plane in three-dimensional space charge solvers.

As coherent dipole modes can easily be damped by active feedback systems, their loss of Landau damping is of less practical concern. The intrinsic damping of quadrupolar coherent bunch oscillations, which are more directly dependent on space charge, and the related intensity thresholds should be a future focus of the studies.
\bibliographystyle{aipnum4-1}   
\bibliography{export.bib}

\end{document}